\documentstyle[eqsecnum,aps,twocolumn]{revtex}
\input epsf.tex
\begin{document}
\draft
\title{On the interpretation of spin-polarized \\
electron energy loss spectra}
\author{R. Saniz}
\address{Departamento de Ciencias Exactas, Universidad Cat\'olica
Boliviana, Casilla \#5381, Cochabamba, Bolivia}
\author{S. P. Apell}
\address{Department of Applied Physics, Chalmers University of
Technology and G\"oteborg University, S-41 296, G\"oteborg, Sweden}
\date{March 3, 2000}
\maketitle
\begin{abstract}
We study the origin of the structure in the spin-polarized electron
energy loss spectroscopy (SPEELS) spectra of ferromagnetic
crystals. Our study is based on a $3d$ tight-binding Fe model, with constant
onsite Coulomb repulsion $U$ between electrons of opposite spin. We find
it is not the total density of Stoner states as a function of energy loss
which determines the response of the system in the Stoner region,
as usually thought, but the
densities of Stoner states for only a few interband transitions.
Which transitions are important depends ultimately on how strongly umklapp
processes couple the corresponding bands.
This allows us to show, in particular,
that the Stoner peak in SPEELS spectra
does not necessarily indicate the value of the exchange splitting energy.
Thus, the common assumption that this peak allows us to
estimate the magnetic moment through its correlation with exchange
splitting should be reconsidered, both in bulk and surface studies.
Furthermore, we are able to show that the above mechanism is one of the main
causes for the typical broadness of experimental spectra.
Finally, our model predicts that optical
spin waves should be excited in SPEELS experiments.
\end{abstract}
\begin{twocolumn}
\narrowtext
\section{Introduction}
\label{intr}
The study of elementary excitations in itinerant-electron ferromagnets
is an area which is currently very active in spite of the enormous amount of
publications on the subject
since early work in the sixties (see e.g. Refs. \onlinecite{kubo62,herring66}
and references therein).
From the beginning, experimental and theoretical work on these materials
concentrated on
neutron scattering and dynamical susceptibility
studies.\cite{kubo62,mook69,mook73,lowde70,cooke73} Efforts
have continued in this direction until today, both because of the
gradual improvement of
electronic band structure calculations\cite{cooke80,tang98} and
because of the improvement of the experimental method.\cite{perring91}
Around
the mid-eighties, however, a new technique was introduced in the field,
namely, spin polarized electron energy loss spectroscopy (SPEELS). Among
its first successes, one can count the first observations, in a ferromagnetic
glass\cite{hopster84} and in nickel,\cite{kirschner84} of what were
interpreted as Stoner excitations.
Further work, reporting more detailed measurements, confirmed those
findings.\cite{venus88,abraham89}
Theoretical model calculations of inelastic electron
spin-flip exchange scattering,\cite{yin81,glazer84,vignale85,bocchetta87}
provided a basis for the interpretation of those observations in terms
of Stoner excitations.
In addition, Vignale and Singwi found in their work
that spin waves should also be observable in SPEELS
measurements.\cite{vignale85} However, these had not been observed at the
time, nor were they observed in the several years that followed.
Spin waves were found in other model calculations,\cite{saniz93,plihal98}
the calculations of Plihal and Mills being the most
conclusive in this respect because of their more accurate treatment
of electronic structure.\cite{plihal98}
It is only very recently that the detection of spin waves in a
SPEELS experiment has finally been reported.\cite{plihal99}
The application of SPEELS has been naturally extended
to the study of magnetic surfaces.\cite{kirschner90,walker92,hopster94}
An important
theoretical effort in this direction is that by Mills and
collaborators,\cite{gokhale92,gokhale94,plihal95} who
have studied ferromagnetic thin films as well.

To introduce the questions addressed by this work, we recall briefly some
of the main concepts involved in SPEELS and discuss some of the findings to
date.
SPEELS is a spin-polarized version of electron energy loss spectroscopy in
the sense that the spin polarization of the scattered electrons is also
measured. The impinging electron often is also spin-polarized, but this is
not necessarily so (see e.g. Refs.
\onlinecite{hopster84,kirschner84}).
In a so-called spin-flip exchange scattering event, an incoming
electron with given spin comes to occupy an empty level in the material
while an electron with opposite spin is driven out and
is detected. The process produces thus a Stoner excitation.
In the band picture of magnetic transition metals, below the Curie
temperature, the exchange
split 3$d$ bands provide large densities of occupied majority-spin states below
the Fermi energy and vacant minority-spin states above it. Thus, it is more
likely for an impinging electron with minority spin to excite a Stoner pair
than for a majority-spin incoming electron, particularly for an
excitation energy corresponding
to the exchange splitting of the ferromagnet. This is the mechanism invoked
to explain the Stoner peak or the asymmetry reported in Refs.
\onlinecite{hopster84} and \onlinecite{kirschner84} and further experimental
work (Refs. \onlinecite{venus88,abraham89}). However,
it turned out necessary to elaborate on several other issues.
Firstly, the Stoner peak was very broad in all observations.
This was interpreted by
Kirschner, Rebenstorff, and Ibach\cite{kirschner84} as an indication
of the nonuniformity of exchange splitting throughout the
Brillouin zone. Then, in Fe, the energy loss at
which the Stoner peak occurs and its width were reported by Venus and Kirschner
to increase with increasing scattering angle,\cite{venus88} a fact that was
correlated by these authors with the calculated
density of Stoner states. Also, a threshold for the onset of
Stoner excitations in Ni(110) was reported by Abraham and Hopster
\cite{abraham89} and interpreted in
terms of the Ni 3$d$ band structure. These workers, moreover,
indicated that their
spectra did not differ significantly for off specular scattering angles
ranging from 10$^\circ$ to 40$^\circ$,\cite{comment}
which they explained as due to
the non-conservation of the momentum component perpendicular to the
surface.

Finally, an important
application based on SPEELS interpretation is that,
in surface and thin film studies, the Stoner peak is assumed
to give information on the surface
magnetic moment through the correlation between exchange splitting and
moment.\cite{walker92,hopster94} In particular, a Stoner peak found
at higher energies than the exchange splitting bulk value is assumed to
indicate an enhanced magnetic moment at the surface.

Clearly, more theoretical work is required, for the bulk as much as for
surfaces, to make further progress. In particular, it would be important
to understand better the phenomenology of SPEELS and to try to be more specific
about the information we can expect from it. This would also provide
experimenters with useful feedback.
Accordingly, we think it
is worthwhile going back to a model calculation and look more closely
at the dynamic properties of the material probed by SPEELS.
In this work we consider a model of Fe based on paramagnetic
tight-binding 3$d$
bands, with up and down spin bands rigidly split. The cross section for
spin-flip exchange scattering processes is evaluated within the random phase
approximation, assuming the solid is described by
a multi-band Hubbard Hamiltonian. As we shall see, this allows
us to show that it is not the total density of Stoner states as a function of
energy loss and momentum transfer which causes the structure in the Stoner
region of the spectrum, but the
density of Stoner states for a few interband excitations. Which interband
excitations are important is essentially determined by the
weight of the matrix elements for such processes.
It this regard the contribution of umklapp scattering is
fundamental because of the
the coupling of different bands at different energy ranges.
This gives rise to a richer
structure in the Stoner region of the spectra.
Also, our model predicts that optical spin waves should be observable
through SPEELS. Again umklapp scattering proves critical, providing optical
spin waves with the necessary oscillator strength. We find this strength
to depend importantly on scattering angle.

Section II of this paper is devoted to theory, presenting the derivation of
the spin-flip exchange
scattering cross section for our model. We present our main results in
Section III. Then follows in Section IV a discussion of our results in the
light of
experimental findings and other theoretical work.
Finally, in Section V we summarize our work and give some conclusions.

\section{Theory}
\subsection{Spin-flip exchange scattering cross section}

The electron spin-flip exchange scattering
differential cross section for a $N$-electron system target
 has been previously derived on general grounds by Vignale and Singwi
 in terms of a particle-hole excitation
correlation function.\cite{vignale85}
One has
\begin{equation}
{{d^2\sigma}\over{dE\,d\Omega}}=-{m^2\over{4\pi^2\hbar^4}}{p_f\over{p_i}}
{1\over\pi}
{{{\rm Im}\,\chi^R_{\sigma_i\sigma_f}({\bf p}_i,{\bf q},E)}\over{1-e^{-\beta E}}},
\end{equation}
where $E$ is the energy loss, $\Omega$ is the solid angle, $m$ the
electron mass, and $\beta=1/k_BT$.
Momentum transfer is given by ${\bf q}={\bf p}_i-{\bf p}_f$, with ${\bf p}_i$
and ${\bf p}_f$ the momentum of the incoming and outgoing electrons,
respectively. Likewise, $\sigma_i$ is the spin of the impinging electron and
$\sigma_f$ that of the scattered one.
The retarded function $\chi^R_{\sigma_i\sigma_f}$ can be obtained by analytic
continuation of the two-particle temperature correlation function
\begin{eqnarray}
\chi_{\sigma_i\sigma_f}({\bf p}_i,{\bf q},i\omega_n)&&=-\int^\beta_0 d\tau\,
e^{-i\omega_n\tau} \nonumber \\
& &\times \langle\, T_\tau[\varrho_{\sigma_i\sigma_f}
({\bf p}_i,{\bf q},\tau)\varrho^\dagger_{\sigma_i\sigma_f}({\bf p}_i,{\bf q})]
\,\rangle.
\label{corrfun}
\end{eqnarray}
In this equation $\omega_n=2\pi n/\beta$ is a bosonic Matsubara frequency,
$T_\tau$ is the imaginary time ordering operator, and the brackets indicate the
thermodynamic average in the canonical ensemble.\cite{fetter71}
$\varrho^\dagger_{\sigma_i\sigma_f}$ is the particle-hole creation operator
\begin{eqnarray}
\varrho^\dagger_{\sigma_i\sigma_f}({\bf p}_i,{\bf q})=
-{1\over N}&&\sum_{j=1}^N\int d{\bf r}\,d{\bf r}_j
e^{-i{\bf p}_f\cdot{\bf r}_j}e^{i{\bf p}_i\cdot{\bf r}} \nonumber \\
&& \times v(|{\bf r}-{\bf r}_j|)
\psi_{\sigma_i}^\dagger({\bf r})\psi_{\sigma_f}({\bf r}_j),
\end{eqnarray}
where $\psi_\sigma^\dagger({\bf r})$ is the field operator creating an
electron of spin $\sigma$ at position ${\bf r}$
and $v(r)=e^2/r$ is the Coulomb interaction between the scattered and target 
electrons. The sum runs over the $N$ electrons in the target system.
This expression is quite general, and could be applied equally well to a
solid, an atom, or a molecule.

We now consider the $N$ electrons in a crystal material. We write
the Bloch wave function for a state with wave vector ${\bf k}$ and spin
$\sigma$ in band $n$ in terms of Wannier functions,
\begin{equation}
\psi_{n{\bf k}\sigma}({\bf r})={1\over\sqrt{N_0}}\sum_{\bf R}
e^{i{\bf k}\cdot{\bf R}}\phi_{n{\bf k}}({\bf r}-{\bf R})\eta_\sigma,
\end{equation}
where $N_0$ is the number of sites in the crystal and $\eta_\sigma$ is the
spin function.
Denoting by $a^\dagger_{n{\bf k}\sigma}$ the operator
creating an electron in such a state, the field operators can be expanded as
$\psi^\dagger_\sigma({\bf r})=\sum_{n{\bf k}}\psi_{n{\bf k}\sigma}({\bf r})
a^\dagger_{n{\bf k}\sigma}$.
The particle-hole creation operator becomes
\begin{equation}
\varrho^\dagger_{\sigma_i\sigma_f}({\bf p}_i,{\bf q})=\sum_{nn'}
\sum_{\bf k}W_{nn'}({\bf p}_i,{\bf q},{\bf k})
a^\dagger_{n{\bf k}\sigma_i}
a^{\phantom{\dagger}}_{n'{\bf k}-{\bf q}\sigma_f},\label{eqph}
\end{equation}
where the sum in momentum space runs over the Brillouin zone and matrix
element $W_{nn'}$ is given by
\begin{eqnarray}
W_{nn'}({\bf p}_i,{\bf q},{\bf k})={{N_0}\over V}\sum_{\bf K}
\hat
v&&({\bf k}-{\bf p}_i-{\bf K})\,\hat\phi^*_{n{\bf k}}({\bf k}-{\bf K})
\nonumber \\ &&\times
\hat\phi_{n'{\bf k}-{\bf q}}({\bf k}-{\bf q}-{\bf K}).\label{eqw}
\end{eqnarray}
Here
${\bf K}$ denotes vectors in the reciprocal lattice and
$V$ is the volume of the sample.
The $\hat{\phantom{o}}$ indicates a Fourier transformed function and $^*$
denotes complex conjugation.
To write the last two equations we have
defined $a^{\phantom{\dagger}}_{n{\bf k}+{\bf K}\sigma}\equiv
a^{\phantom{\dagger}}_{n{\bf k}\sigma}$ and
have exploited the periodicity of the Wannier functions
in the wave vector index, i.e.
$\phi_{n{\bf k}+{\bf K}\sigma}=\phi_{n{\bf k}\sigma}$.

\subsection{RPA expression for a tight-binding system}

We are interested in the cross section for an itinerant electron
ferromagnet. We describe the system within a tight-binding approximation,
thus writing
the Wannier wave function for given ${\bf k}$ and band index $n$
by a linear combination of atomic orbitals $\varphi_m$
\begin{equation}
\phi_{n{\bf k}}({\bf r})=\sum_mb_{mn}({\bf k})\varphi_m({\bf r}).
\end{equation}
The coefficients $b_{mn}$ diagonalize the crystal Hamiltonian
and are normalized so as to define a unitary matrix. Consequently, the
independent-electron Hamiltonian of the system can be written
$H_0=\sum_{n{\bf p}\sigma}\epsilon_n({\bf p})a^\dagger_{n{\bf p}\sigma}
a^{\phantom{\dagger}}_{n{\bf p}\sigma}$, where the $\epsilon_n({\bf p})$
are paramagnetic band energies. We assume the interacting system
is described by a multi-band Hubbard Hamiltonian.
In our basis we have
\begin{eqnarray}
H_I={1\over 2}
{U\over{N_0}}\sum_{{nn'}\atop{mm'}}&&\sum_\sigma\sum_{{\bf p}{\bf p}'{\bf q}'}
c_{nm}({\bf p}+{\bf q}',{\bf p})\,c_{n'm'}({\bf p}'-{\bf q}',{\bf p}')
\nonumber \\
&&\times a^\dagger_{n{\bf p}+{\bf q}'\sigma}
a_{m{\bf p}\sigma}
a^\dagger_{n'{\bf p}'-{\bf q}'-\sigma}
a^{\phantom{\dagger}}_{m'{\bf p}'-\sigma},
\end{eqnarray}
where we have defined
$c_{nm}({\bf p},{\bf q})=\sum_l b_{ln}({\bf p})b_{lm}({\bf q}),$ and
$U$ is the effective on-site Coulomb interaction for two electrons with
opposite spin.
The RPA evaluation of the correlation function $\chi_{\sigma_i\sigma_f}$
defined in Eq.~\ref{corrfun} is a straightforward
generalization of that in previous work \cite{saniz93}.
The response function divides naturally in two,
\begin{equation}
\chi_{\sigma_i\sigma_f}({\bf p}_i,{\bf q},i\omega_n)=
\chi_{\sigma_i\sigma_f}^{\rm S}({\bf p}_i,{\bf q},i\omega_n)+
\chi_{\sigma_i\sigma_f}^{\rm MB}({\bf p}_i,{\bf q},i\omega_n).
\end{equation}
The Stoner or single-particle excitation contribution is given by
\begin{eqnarray}
\chi_{\sigma_i\sigma_f}^{\rm S}({\bf p}_i,{\bf q},i\omega_n)=
\sum_{nn'}\sum_{\bf k}&&
{{f_{n'{\bf k}-{\bf q}\sigma_f}-f_{n{\bf k}\sigma_i}}\over
{i\omega_n+\epsilon_{n'\sigma_f}({\bf k}-{\bf q})-\epsilon_{n\sigma_i}({\bf k})}} \nonumber \\
&&\times |W_{nn'}({\bf p}_i,{\bf q},{\bf k})|^2.\label{eqstoner}
\end{eqnarray}
We have introduced the occupation
probability of state $n{\bf k}\sigma$,
$f_{n{\bf k}\sigma}=\langle a^\dagger_{n{\bf k}\sigma}
a_{n{\bf k}\sigma}\rangle$,
and the single-particle energy modified by the exchange
self-energy
\begin{equation}
\epsilon_{n\sigma}({\bf k})=\epsilon_n({\bf k})-{U\over{N_0}}\sum_{\bf p}
f_{n{\bf p}\sigma}.
\end{equation}
Thus, in this model, spin down and spin up energy bands are rigidly split by 
the quantity
$\Delta=U(\langle n_\uparrow\rangle-\langle n_\downarrow\rangle),$
where
\begin{equation}
\langle n_\sigma\rangle={1\over{N_0}}\sum_{m{\bf p}}f_{m{\bf p}\sigma}
\end{equation}
is the average number per site of states with spin $\sigma$.

The many-body contribution is given by
\begin{eqnarray}
\chi_{\sigma_i\sigma_f}^{\rm MB}({\bf p}_i,{\bf q},i\omega_n)=
{U\over{N_0}}\sum_{nn'}&&G^{nn'}_{\sigma_i\sigma_f}
({\bf p}_i,{\bf q},i\omega_n)
\nonumber \\
&&\times \Gamma^{nn'}_{\sigma_i\sigma_f}({\bf p}_i,{\bf q},i\omega_n),
\end{eqnarray}
with the auxiliary functions $G^{nn'}$ and $\Gamma^{nn'}$ defined as follows:
\begin{eqnarray}
G^{nn'}_{\sigma_i\sigma_f}\!({\bf p}_i,{\bf q},&&i\omega_n)=
\sum_{mm'}\sum_{\bf k}\!
{{f_{m'{\bf k}-{\bf q}\sigma_f}-f_{m{\bf k}\sigma_i}}\over
{i\omega_n+\epsilon_{m'\sigma_f}
({\bf k}-{\bf q})-\epsilon_{m\sigma_i}({\bf k})}} \nonumber \\ && \times
W^*_{mm'}({\bf p}_i,{\bf q},{\bf k})b_{nm}({\bf k})b_{n'm'}({\bf k}-{\bf q}),
\end{eqnarray}
and, considering $G$ and $\Gamma$ as vectors with coefficients indexed by
$nn'$,
\begin{eqnarray}
\Gamma_{\sigma_i\sigma_f}({\bf p}_i,{\bf q},i\omega_n)=&&
\left[ 1+{U\over{N_0}}
D_{\sigma_i\sigma_f}({\bf q},i\omega_n) \right]^{-1} \nonumber \\
&& \phantom{aaa}
\times G_{\sigma_i\sigma_f}({\bf p}_i,{\bf q},i\omega_n),
\end{eqnarray}
where the elements of matrix $D$ are
\begin{eqnarray}
&&D^{nn',mm'}_{\sigma_i\sigma_f}({\bf q},i\omega_n)=\sum_{ll'}\sum_{\bf k}
b_{nl}({\bf k})b_{n'l'}({\bf k}-{\bf q}) \nonumber \\
&&\times
{{f_{l'{\bf k}-{\bf q}\sigma_f}-f_{l{\bf k}\sigma_i}}\over
{i\omega_n+\epsilon_{l'\sigma_f}({\bf k}-{\bf q})-\epsilon_{l\sigma_i}({\bf k})}}
b_{ml}({\bf k})b_{m'l'}({\bf k}-{\bf q}).
\end{eqnarray}
To obtain $\chi^R_{\sigma_i\sigma_f}$, analytic continuation
$i\omega_n\to E+i\eta$ of the above results is straightforward. Also,
results in next section are
obtained in the zero-temperature limit.

\section{Results}

Before presenting our results, there are a few details of our model we
should discuss. As we said in the introduction, our Fe model is based
on simple $3d$ tight-binding paramagnetic bands, i.e.
we neglect $sp$ hybridization. This has, of course, incidence on
the quantitative details of the response of the system.
However, we will see our results are rather general and do not
depend on these details.
The Wannier functions were
written as a linear combination of the Fe five 3$d$ atomic wave functions.
The overlap integrals were
calculated, up to next nearest neighbors, using the Fe atomic wave
functions determined according to Griffith's prescription,\cite{griffith61}
and a lattice constant $a=2.87$ \AA.\cite{ashcroft76} We have, thus, a
five band model. The bandwidth was set to 4.7 eV,
which corresponds roughly to the bandwidth of $d$ electrons
in Fe.\cite{saniz98}
Exchange splitting $\Delta$ was chosen to be 2 eV, taking as reference
the position of the peaks in the densities of states for up and down spins.
Then, the Fermi level was fixed
by the condition of having six electrons per unit
cell. We show the density of states for up and down spins in Fig.~\ref{fig1}.
These exhibit the bonding and antibonding regions common to BCC materials
with unfilled $d$ shells. Once the Fermi energy is fixed we can deduce
the strength of the effective Coulomb interaction $U$ in our model from 
$U=\Delta/(\langle n_\uparrow\rangle-\langle n_\downarrow\rangle).$
We find $U=0.69$ eV, which compares very well with the energy found in
other works.\cite{wohlfarth80} On the other hand, bulk polarization
is too high, roughly 48\%,\cite{comment1} reflecting the lack of
hybridization with $sp$ electrons.
\vspace{0.3cm}
\begin{figure}
\epsfxsize=.95\hsize
\epsfbox{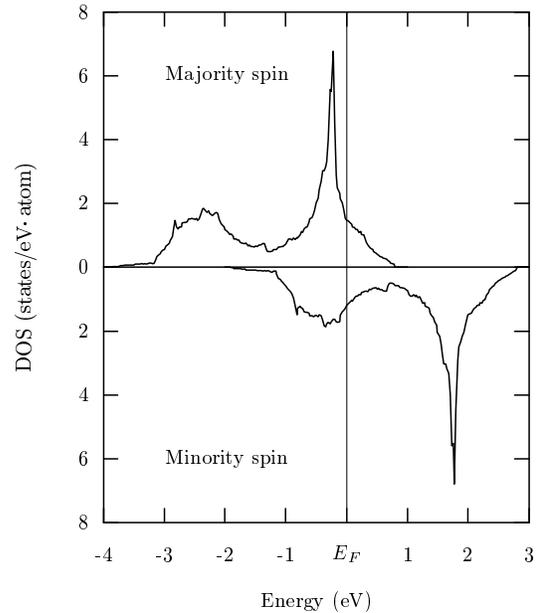}
%\begin{figure}
\vspace{.2cm}
\caption{The Fe up and down spin electron densities of states for
our model, exhibiting the characteristic bonding and antibonding regions.
Bandwidth is 4.7 eV and exchange splitting 2 eV.}
\label{fig1}
\end{figure}

In a typical SPEELS experiment,
the incoming electron beam impinges on the sample surface at
an angle $\theta$ to the normal and the total scattering angle is
90$^\circ$.\cite{kirschner84,venus88,abraham89,walker92,hopster94}
For fixed scattering
angle, i.e. for given incoming and outgoing momenta, there are three
possible scattering processes corresponding to different momentum transfer.
In two cases, a relatively small angle inelastic scattering event is
preceded or followed by elastic scattering.
In the third one, all the momentum transfer is absorbed by the electron-hole
pair excitations, i.e. it is a large angle scattering event.
We consider here
the geometry of Venus and Kirschner,\cite{venus88} that is, the sample
exposes the (110) surface and the scattering plane is defined by the
surface normal [110] and the [001] axis. If $u$ is the axis
normal to the surface, the three momenta mentioned are given as a function
of energy loss $E$ and impinging momentum $p_i$ and energy $E_i$ by
\begin{eqnarray}
& &q_u=p_i(\cos\theta-\sin\theta\sqrt{1-E/E_i}), \nonumber \\
& &q'_u=p_i(-\cos\theta+\sin\theta\sqrt{1-E/E_i}), \nonumber \\
& &q''_u=-p_i(\cos\theta+\sin\theta\sqrt{1-E/E_i}), \nonumber \\
& &q_z=q'_z=q''_z=p_i(\sin\theta-\cos\theta\sqrt{1-E/E_i}),\label{eqq}
\end{eqnarray}
where $\theta$ is the angle to the normal. The large angle scattering event
corresponds to ${\bf q}''$.
Momentum transfer parallel to the surface is the same in the three
cases. To calculate the final spectrum, the contributions of these three
processes have to be added because experiment does not discriminate them.

Also, since Fe presents a non negligible quantity of
free-like $s$ and $p$ states at the Fermi surface,
the interaction between the incoming electrons
and those in the solid will be screened. We take this into account using
the Thomas-Fermi form of the
screened Coulomb potential, with a
screening wave vector corresponding to the density
of states of $s$ and $p$ electrons at the Fermi surface.\cite{sukjoo99}
This gives $q_{\rm TF}=0.26$ in units of
$k_a=4\pi/a$. Finally, an important point to mention is that
we use a finite value for $\eta$ when taking the analytic continuation
$i\omega_n\to E+i\eta$. Since, to our knowledge, there are no estimates of
the self-energy corrections for up and down spin bands in
Fe,\cite{starnberg88} we take $\eta=80$ meV,
which corresponds to the resolution in the latest experiment on this material.
\cite{plihal99}

\subsection{Interband densities of Stoner states}
\label{sdos}

Let us consider a majority spin electron with
an angle of incidence $\theta=60^\circ$ to the normal
and an incoming energy $E_i=22$ eV,
which is the energy used in Ref.~\onlinecite{venus88}.
We see in Fig.~\ref{fig2}(a) that the total spin-flip exchange scattering
cross section is indeed rather broad,
with its peak centered at an energy much higher than the exchange
splitting value (2 eV in our model), a trend observed experimentally by
Venus and Kirschner\cite{venus88}.
We also show the partial cross sections for different momentum transfers.
The curves for small scattering angle coincide for
\begin{figure}
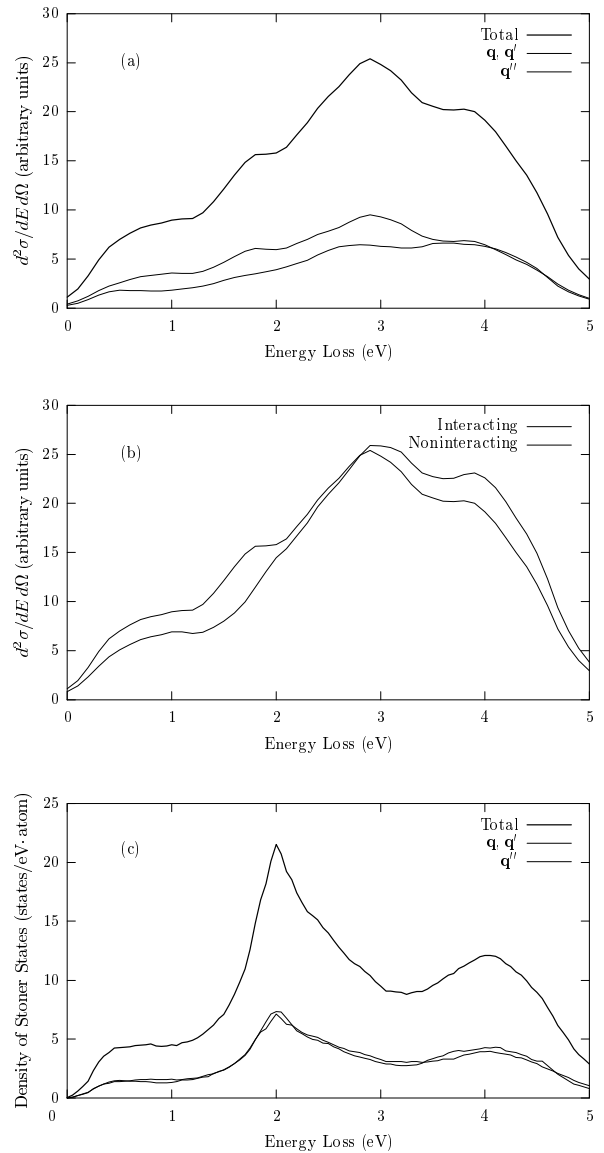
\null\vspace{-25pt}
\epsfxsize=1.045\hsize
\begin{flushright}\epsfbox{fig2aeps}\end{flushright}
\epsfxsize=1.045\hsize
\null\vspace{-30pt}
\begin{flushright}\epsfbox{fig2beps}\end{flushright}
\epsfxsize=1.045\hsize
%\vspace{0.04truecm}
\begin{flushright}\epsfbox{fig2ceps}\end{flushright}
%\begin{figure}
\caption{(a) Spin-flip exchange cross section for a majority spin
electron, with impinging energy of 22 eV and
angle of incidence of 60$^\circ$. We show the total cross section, as well
as the partial cross sections for different momentum transfer.
The peak is at 3 eV, an energy
much higher than the exchange splitting $\Delta=2$ eV.
(b) Interacting and noninteracting cross sections. The difference between
both curves clearly shows two collective modes, one just below 2 eV, and
the other below 1 eV. The noninteracting cross section shows three distinct
features, namely, the peak at 3 eV, a shoulder at 4 eV, and a broad hump
around 1 eV.
(c) The total and partial densities of Stoner states. These show the typical
maxima at exchange splitting, which are absent from the SPEELS spectrum.
The densities of Stoner states are incapable of explaining the peak of
the SPEELS spectrum at 3 eV.}
\label{fig2}
\end{figure}
\noindent symmetry reasons.
Though total cross section broadness is somewhat
increased because of the difference between small angle and large angle
scattering, the cross section in each case is broad in itself.
We have examined the origin of the structure in this spectrum.
Firstly, we separated single-particle
excitations and many-body effects. In Fig.~\ref{fig2}(b) we show the
noninteracting and interacting (total) cross sections. This figure clearly
shows us the contributions of collective modes. Indeed, the broad feature
starting around 0.2 eV indicates the excitation of low lying spin waves,
and, more interestingly, the shoulder at higher energy, below 2 eV,
indicates the excitation of optical spin waves.
This is important because optical spin waves have
not been discussed previously in connection with SPEELS measurements.
We consider spin waves again further on and we concentrate
here on the single particle traits.
Besides the peak at 3 eV, the noninteracting cross section shows a
shoulder at 4 eV and a broad feature, albeit much smaller,
at low energy loss, around 1 eV or so. SPEELS spectra
have often been interpreted in terms of the density of Stoner
states. Accordingly, we show in Fig.~\ref{fig2}(c) the total density of
Stoner states, as well as the densities of Stoner states for different
momentum transfer (as before, the curves for small angle scattering are
the same).\cite{comment5}
It is evident in the cross sections that
there is nothing reminiscent of the high density of states at the exchange
splitting energy. The only features of the cross sections that can find
an explanation
in the density of Stoner states are the shoulder at 4 eV and, possibly,
the hump around 1 eV.
We have, thus, refined our study and have considered the behavior of the
density of Stoner states as a function of energy loss and the bands coupled
in an excitation (recall energy loss and momentum transfer are coupled, cf.
Eq.~(\ref{eqq}))
\begin{eqnarray}
\rho_{nn'}(E)
={1\over{N_0}}&&\sum_{\bf k}
(f_{n'{\bf k}-{\bf q}\sigma_f}-f_{n{\bf k}\sigma_i}) \nonumber \\ &&\times
\delta(E+\epsilon_{n'\sigma_f}({\bf k}-{\bf q})-\epsilon_{n\sigma_i}({\bf k})).
\end{eqnarray} 
Hence, subscripts $n$ and $n'$ indicate minority and majority bands,
respectively (bands are numbered from bottom to top).
We show a plot of the density of states thus defined in Fig.~\ref{fig3}(a).
We can see that the allowed and forbidden
interband excitations are completely identified (we use the term forbidden
for interband excitations with vanishing density of Stoner states).
Moreover, the series of Stoner peaks clearly reflect the bonding and
antibonding nature of the electronic structure, giving rise to two arrays
of peaks, for higher and lower excitation energies.\cite{comment2}
The question is, of course, which of these Stoner peaks contribute
the most to SPEELS cross sections. The answer is to be found, perhaps
unsurprisingly, in how strongly the different bands are coupled
by the matrix elements $W_{nn'}$ of the electron-hole creation operator,
which weights the contribution of each Stoner excitation
(see Eq.~\ref{eqstoner}).
What is not so obvious is the outcome of the combined effect of Stoner
peaks and matrix elements.
Let us consider the average
of the square of the absolute value of matrix elements
$W_{nn'}$ over the Brillouin zone, i.e.
\begin{figure}
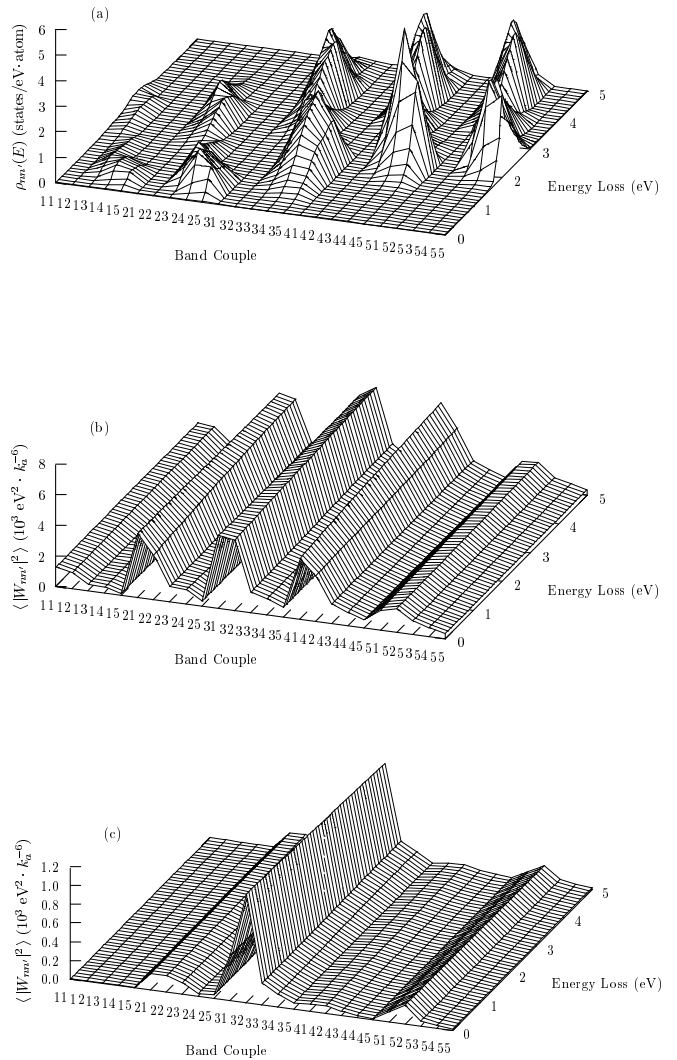

\epsfxsize=1.1\hsize
\begin{flushright}\epsfbox{fig3aeps}\end{flushright}
\epsfxsize=1.1\hsize\null\vspace{5pt}
\begin{flushright}\epsfbox{fig3beps}\end{flushright}
\epsfxsize=1.08\hsize
\vspace{0.04truecm}\null\vspace{5pt}
\begin{flushright}\epsfbox{fig3ceps}\end{flushright}
%\begin{figure}
\caption{(a) 3D plot of the interband densities of Stoner states. Allowed and
forbidden interband excitations are clearly identified. The two series of
peaks reflect the bonding and antibonding nature of the Fe electronic
structure. The highest peak is found at exchange splitting, for $nn'=44$.
As we explain in the text, however, this peak contributes little to the
spin-flip exchange cross section.
(b) 3D plot of the average $\langle\,|W_{nn'}|^2\,\rangle$.
The slight dependence
on energy results in the wave like form of the surface.
The lines on the surface parallel to
the energy axis correspond to fixed band couple value.
The important interband
excitations are determined by the crests, namely 21,22,31,32,41, and 42.
One can also observe that $\langle\,|W_{nn'}|^2\,\rangle$
reaches its lowest values for $n5$, with $n=1,\ldots,5$.
(c) 3D plot of
$\langle\,|W_{nm}|^2\,\rangle$ without umklapp processes,
The important interband
excitations have been reduced to $nn'=31,32$, thus singling out excitations
in a restricted energy range.
}
\label{fig3}
\end{figure}
\vspace{5pt}
\begin{equation}
\langle\,|W_{nn'}|^2\,\rangle={1\over{\hat v}}\sum_{\bf k}|W_{nn'}|^2
\end{equation}
($\hat v$ denoting the volume of the Brillouin zone).
In Fig.~\ref{fig3}(b) we show the graph of
$\langle\, |W_{nn'}|^2\,\rangle$ as a
function of energy loss and of bands coupled. There is little significant
variation as a function of energy loss, but a very important
structure as a function of band couple, resulting in a wave like
pattern. Comparing Figs.~\ref{fig3}(a) and \ref{fig3}(b) we can
clearly see when it
is that both quantities, $\,\rho_{nn'}$
and
$\langle\,|W_{nn'}|^2\,\rangle$, interfere
constructively.
Thus, although the density of Stoner states reaches is
highest peak
at exchange splitting, the average
$\langle\,|W_{nn'}|^2\,\rangle$ is negligible
for the corresponding band couples.
Instead, although the densities of Stoner states for
interband excitations 21 and 22, 31 and 32, and 41 and 42 
are more modest, the corresponding matrix element averages are high.
Whence the 
peak around 3 eV and the shoulder around 4 eV in the
noninteracting cross section in Fig.~\ref{fig2}(b). Actually, the peak at
3 eV is more of a hat on top of the high cross section value due to
interband excitations 31 and 32.
We can also see that the hump around 1 eV
is due to excitations coupling bands 2 and 3, and 2 and 4.
To corroborate our analysis, we show
in Fig.~\ref{fig4} the cross section taking into account solely
the interband processes mentioned above. We include the total noninteracting
cross section for comparison as well.
We see that the few interband excitations considered indeed
account almost completely for the structure of the noninteracting spectrum. 
An argument to understand how so simple a picture can work
is that, since the atomic 3$d$ orbitals are localized, their
Fourier transform is rather flat, so that $|W_{nn'}|^2$ in the single-particle
correlation function
$\chi_{\sigma_i\sigma_f}^{\rm S}$ (cf. Eq.~(\ref{eqstoner}))
may be replaced by its average value over the Brillouin zone. Thus,
$\chi_{\sigma_i\sigma_f}^{\rm S}$
is approximately proportional to
$\sum_{nn'}\rho_{nn'}\langle |W_{nn'}|^2\rangle$,
a weighted average of the interband densities of Stoner states.

\null\vspace{10pt}
\begin{figure}
\epsfxsize=1.1\hsize
\epsfbox{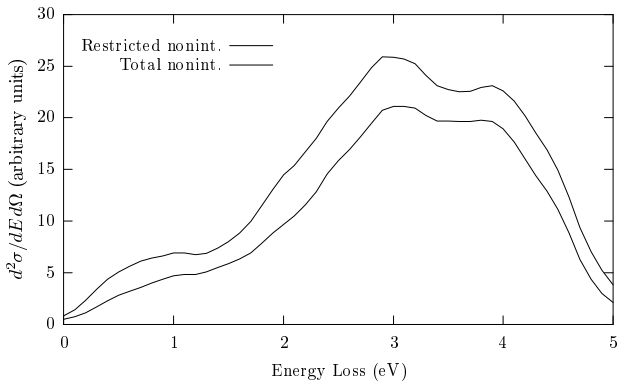}
%\begin{figure}
\vspace{5pt}
\caption{Noninteracting scattering cross section taking into account only
8 ($nn'=21,22,23,24,31,32,41,42$)
out of the 25 possible interband excitations, selected as explained
in the text, compared to the total noninteracting scattering cross section.
The first curve
follows very closely to the second one.}
\label{fig4}
\end{figure}

\begin{figure}
\epsfxsize=1.1\hsize
\null\vspace{-5pt}\epsfbox{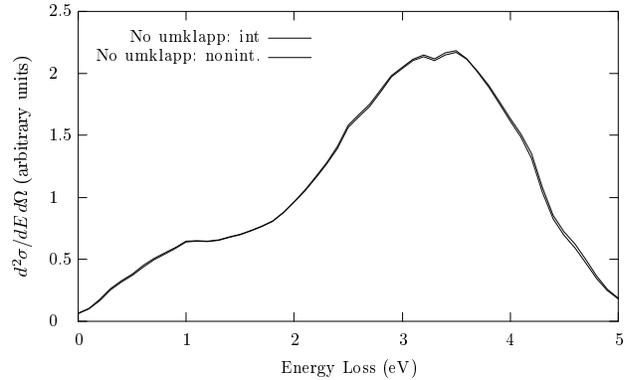}
%\begin{figure}
\vspace{5pt}
%\begin{figure}
\caption{Spin-flip exchange scattering cross section without umklapp processes.
All that is left is a broad maximum around 3 and 4 eV, and the small
hump at small energy loss. Clearly, the most
important information is lost, i.e. the peak at 3 eV and the
shoulder at 4 eV (cf.~Fig.~\ref{fig2}(b).)}
\label{fig5}
\end{figure}

\subsection{Umklapp processes}

Another most interesting phenomenon playing a fundamental role in
SPEELS is umklapp scattering.
One can see in Eq.~(\ref{eqw}) that the contribution of umklapp processes 
to the particle-hole excitation operator $\varrho_{\sigma_i\sigma_f}$ is
weighted by
the Coulomb interaction and the Wannier
wave functions.
Because of the decay of the Coulomb potential as
well as of the atomic orbitals with increasing wave vector,
the weight becomes rapidly
negligible for lattice vectors beyond first nearest neighbors.
This is enough, however, for umklapp processes to have a twofold
effect. To see this, let us consider cross sections
taking into account only normal excitations (i.e. with
respect to the first Brillouin zone). We show this in Fig.~\ref{fig5},
where we plot both the interacting and noninteracting cross sections.
Firstly, we see that, quite apart from their much lower values
in comparison with the full response case, spectra in Fig.~\ref{fig5}
show little resemblance with those
in Fig.~\ref{fig2}(b) (scale in both figures is the same).
This is because the possible interband excitations have been drastically
reduced. Indeed, let us consider the graph of the average
$\langle\,|W_{nn'}|^2\,\rangle$ for excitations strictly conserving
crystal momentum.
We show this in Fig.~\ref{fig3}(c).
The wave crests have been reduced to that for $nn'=31,32$,
from which it is obvious
the the different wave crests in Fig.~\ref{fig3}(b) are due to umklapp
scattering. Normal scattering alone
results in a spectrum almost completely distorted because of
the excitation of interband transitions mainly for energies between
3 and 4 eV (cf. Fig.~\ref{fig3}(a)).

Also, in Fig.~\ref{fig5} it is immediately apparent that there
remains no trace of spin waves in the spectrum.
Indeed, the interacting and noninteracting
curves are almost indistinguishable, with no hint of the spin wave
modes below 2 or 1 eV. Thus, it appears that
umklapp processes provide
collective excitations with oscillator
strength, at
\begin{figure}
\epsfxsize=1.05\hsize
\null\vspace{0pt}
\epsfbox{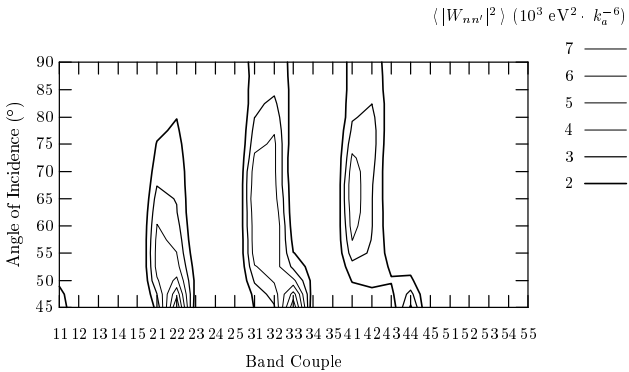}
%\begin{figure}
\null\vspace{-5pt}
\caption{Contour plot of $\langle\,|W_{nn'}|^2\,\rangle$ as a function of
scattering angle,
for an energy loss $E=2.5$ eV. Thinner line correspond to higher averages.
Drastic changes occur particularly
for near specular
scattering. In this case, the interband excitations
for $nn'=22,33$ and 44 will overwhelmingly dominate the spectrum.
Averages decrease significantly for large scattering angles.
Comparing values for $60^\circ$ and
for $55^\circ$, we see that
interband excitations for $nn'=41,42$ are less important for the
latter than for the former. Similarly, for $70^\circ$, interband
excitations 21 and 22 play a less important role than for $60^\circ$.
}
\label{fig6}
\end{figure}
\noindent
least optical modes.
We must point out here that the
low lying collective mode contributing to
the broad feature below 1 eV is
not an acoustic mode, but also an optical one. We have calculated
its dispersion relation in the [100] direction and found that it
tends linearly to 218.7 meV for $q\to 0$. The slope is positive, but
very low, with a value of 8.9 meV$\cdot\,$\AA. 
The reason is that the
energy bands in our model are purely $d$.
It is well known that models of
itinerant ferromagnetism which do not take into account hydridization
with $sp$ bands, fail to describe spin waves
appropriately.\cite{muniz85}
So we do not expect our model to predict accurate dispersion relations
for collective excitations. However, we do think our result properly
introduces optical spin waves as a source of structure in SPEELS
measurements.

\subsection{Spectra and angle of incidence}

To illustrate further the pertinence of our analysis, let us consider an
example.
A question addressed in the past by experimenters, without finding a
clear answer,\cite{venus88,abraham89} has been that of the
variation of spectra with scattering angle. Since
$\langle\,|W_{nn'}|^2\,\rangle$
plays such a consequential role, we take a look at its dependence
on angle of incidence through the contour plot in Fig.~\ref{fig6}.
Generally speaking, the same interband averages remain the most important
important as angle changes, except
toward specular scattering, when weights raise and shift significantly
(then the highest values are to be found for $nn'=22,33$, and 44).
Let us consider the spectra for angles of incidence $\theta=55^\circ$
and $\theta=70^\circ$. In the first case, Fig.~\ref{fig6} shows us that
the importance of interband excitations $nn'=41,42$ is diminished with
respect to the $\theta=60^\circ$ case. For $\theta=70^\circ$,
it is band couples $nn'=21,22$ that are diminished. In this way, in fact, we
hinder the contribution of umklapp processes coupling different
bands. Regarding the densities of Stoner states,
on the other hand, we found that
changes from one angle to another are rather small, affecting
essentially only the height of the peaks.
This is because a change in angle
will not change the energies at which there can be a Stoner
excitation, but basically the number of these. Thus,
considering the averages $\langle\,|W_{nn'}|^2\,\rangle$,
the scattering cross section for $55^\circ$ should present
almost no shoulder around 4 eV,
since interband excitations 41 and 42 are
weak for that angle.
Likewise, we expect the scattering cross section
for $70^\circ$ to have a weaker peak around 3 eV, because of the
negligible contribution of excitations for $nn'=21,22$. We can appreciate
these effects in Fig.~\ref{fig7}. Thus, given the peaks of the interband
densities of Stoner states, the shift in the Stoner excitation
maximum with varying scattering angle
obeys to which interband excitations receive more weight, according to
the corresponding matrix elements averages. This, in turn, depends
on which umklapp processes gain more importance.
Fig.~\ref{fig7} also shows
that the maximum shift causes the broadness of the spectrum to increase
with scattering angle.

Turning to spin waves,
it is interesting to note that the strength of the higher optical
mode decreases with scattering angle. Thus it appears that umklapp
scattering is unable to transmit to it sufficient oscillator strength
at higher scattering angles. On the other hand, through
comparison with the noninteracting cross sections we found that the lower
lying spin wave seems less affected by scattering angle.
\vspace{10pt}
\begin{figure}
\epsfxsize=1.05\hsize
\epsfbox{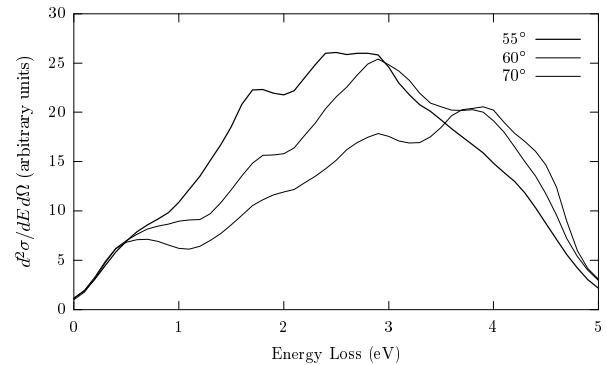}
%\begin{figure}
\vspace{2pt}
\caption{Total cross sections for angles of incidence
$55^\circ$, $60^\circ$ and $70^\circ$.
The main peak shifts toward higher energies according to the weight the
different interband excitations gain or lose because of umklapp
processes and according to which Stoner
peaks are approached. Broadness increases with peak shift.
Also, the higher optical mode clearly
loses strength with increasing scattering angle, while the lower lying
mode seems less affected.}
\label{fig7}
\end{figure}

\section{Discussion}

We wish to discuss some of the issues considered in this work pertaining to
other theoretical and experimental findings.
First of all, as we have shown (cf. Fig.~\ref{fig2}(b)), the maximum in the
SPEELS spectrum does not necessarily correspond to the exchange splitting
energy of the ferromagnet, since the peak is at 3 eV and $\Delta=2$ eV
in our model.
Thus, the common assumption that this peak allows us to
estimate the magnetic moment\cite{walker92,hopster94} through its
correlation with exchange
splitting should be reconsidered, both in bulk and surface
studies.

Also, the broadness of the spectrum is generally
associated with a non constant exchange splitting over the Brillouin zone.
While we agree a non constant exchange splitting will have this effect,
we have seen that a most important source of broadness is umklapp
scattering, together with the structure of the interband
densities of Stoner states of the material.
In particular, even for specular or near specular
scattering, will a model
with rigidly split bands present a relatively broad spectrum. Still,
according to our interband
densities of Stoner states plot, for rigidly split bands we expect to
see a maximum at exchange splitting. Unfortunately there are no reliable
results for specular scattering in the case of Fe.\cite{comment3}
In this regard, the case of Ni would be very interesting to investigate
in more detail. To begin with, the results of Kirschner,
Rebenstorff, and Ibach\cite{kirschner84} and of Abraham and
Hopster\cite{abraham89} are contradictory. Indeed, the former reported a
broad maximum
around exchange splitting for near specular scattering, while the latter
stated that they see no sharp feature at that energy. According to our
picture, Abraham and Hopster's result would be explained only if the matrix
elements $W_{nn'}$ are always weak for energies near exchange splitting.
An accurate calculation of the interband densities of Stoner states
and of the weights $W_{nn'}$ would be most clarifying in this respect.
For off specular scattering, however,
both groups reported a weak dependence on scattering angle.
This is plausible, according to our results. In Fig.~\ref{fig6}
we see that the $\langle |W_{nn'}|^2\rangle$ may remain relatively
unchanged for certain scattering angle intervals, with, consequently,
little change in the SPEELS spectra.

As mentioned in the introduction, moreover, Abraham and Hopster
interpret the onset of Stoner excitations found in their work in
terms of the 3$d$ band structure of Ni. Again, this could readily be verified
having at hand the interband densities of Stoner states for this material.
Indeed, these authors consider in particular interband excitations
corresponding to $nn'=55$, the onset of which, if they exist, could
be easily identified in a graph like that in Fig.~\ref{fig3}(a).
A point still to be verified would be if
the matrix elements for such excitations are sufficiently important.

Spin waves, both acoustic and optical, have long been predicted in
itinerant ferromagnets and subsequently observed through neutron
scattering.\cite{kubo62,mook73,perring91} The question is, then, if
these are observable in
SPEELS. The question to be studied in the future is if
there can be enough coupling between the incoming electron and those
in the solid to excite an optical spin wave. In our model, it is
umklapp processes that provide optical spin waves with the necessary oscillator
strength to contribute significantly to the spectra. It could be,
however, the acoustic waves, when present, drain most of the oscillator
strength. This is plausible because it is known that acoustic spin
waves in Fe arise upon hybridization of $d$ electrons with $sp$ electrons.
Matrix elements with $sp$ hybridization will be more important because
of the larger $s$ wave functions. This implies, of course, as our model does,
that optical modes are mainly $d$ in character.
The case of Ni appears
again to be different, since even a pure $d$ band model of Ni shows
acoustic spin waves.\cite{lowde70}

The analysis presented in this work can prove useful more broadly in the
understanding of ferromagnetism. Recently, Hirsch has presented a model
of ferromagnetism without exchange splitting, in which spin polarization
arises upon broadening of the spin-up bands relative to
spin-down.\cite{hirsch99} If this mechanism plays an important role in
itinerant ferromagnets, then the interband densities of Stoner states
will change considerably
because other pairs of bands will be involved than those in the Stoner
picture of ferromagnetism. Consequently, the predicted exchange scattering
spectra will be different in both pictures. Thus SPEELS can prove a useful
tool to validate or disprove Hirsch's model, possibly improving
resolution previously.

Finally, we would like to comment on the bulk {\it vs.} surface
question.
Our calculations here have focused on bulk properties. However, some authors
have presented SPEELS as a technique more appropriate for surface
studies.\cite{hopster94,gokhale94} The reason is concern regarding
the mean free path of electrons at the energies used in SPEELS.
The question raised, however, is not simple and requires more detailed
consideration, both theoretically and experimentally.
Still, most of
the observations to date have been discussed in the light of bulk
calculations.\cite{venus88,abraham89,plihal99} 
In this regard, a recent report on the electron dynamics at the surfaces
of noble metals (Ag and Cu) is appropriate to mention.
B\"urgi {\it et al.} \cite{burgi99} have
found that the dynamics of hot electrons at surfaces can be dominated by
bulk electrons.
This offers more support to the premise that our results offer a sensible
explanation for SPEELS results.

\section{Summary and conclusions}

In this work we address the problem of the interpretation of the SPEELS
spectrum of itinerant ferromagnets.
We find that considerably more information can be drawn from these
measurements than has been recognized until now. We have found
that the peaks of the spectra in the Stoner
region are the image of a few
interband densities of Stoner states of the material. These are very sensitive
to the electronic structure of the material and illustrate very
clearly the allowed and forbidden interband excitations.
The important band
couples in SPEELS are determined by the average
weight of the squared matrix elements of the
pendant Stoner excitations. In this respect, umklapp processes play a most
fundamental role.
Our model also predicts that optical spin waves should be excited in
SPEELS experiments,
with umklapp scattering providing the
necessary oscillator strength. Our results allow us also to explain several
of the features observed in SPEELS spectra.
From the theoretical point of view,
{\it ab initio} calculations of the interband densities of Stoner states and
matrix elements $W_{nn'}$ would provide closer look to the details of
the mechanisms behind SPEELS.
The differences between
different ferromagnets, like Fe and Ni, could also be better understood.
From the experimental point of view, measurements with higher resolution
would be desirable, particularly for the detection of spin waves.
We think our results provide a good starting point for those
further studies.

\acknowledgments

R.S. would like to thank the members of the
Department of Applied Physics at Chalmers University of Technology for their
hospitality during a stay in which part of this work was done.
This project was supported by the Swedish Natural Science
Research Council. R.S. would also
like to thank Suk Joo Youn and Ove Jepsen for LMTO-ASA data for Fe,
and Anthony Paxton for providing him with useful literature.
Many thanks in particular to Prof. A. J. Freeman for encouraging discussions.

\end{twocolumn}


\begin{references}
\bibitem{comment} A similar claim was made in Ref.~\onlinecite{kirschner84}.
\bibitem{comment5} Note that the densities of Stoner states do not quite
vanish for zero energy loss. For off specular scattering, $E=0$ corresponds
to elastic scattering with finite momentum transfer, which presents a low but
finite density of Stoner states.
\bibitem{comment2} Indeed, it is not difficult to see that
for $n=3,4,5$ the Stoner peaks correspond essentially
to excitations coupling (minority) antibonding holes to (majority)
antibonding electrons, for lower energy cases, and to
bonding electrons for higher energy. The same is
true for the coupling of the $n=1,2$ bonding holes with electrons.
\bibitem{comment3} The measurements of Venus and Kirschner in
Ref.~\onlinecite{venus88} show no peak around exchange splitting for
specular scattering, but the authors warn their results in
that case are not reliable.
\end{references}
\end{document}